\documentclass[aps,pre,epsfig,twocolumn]{revtex4-1}

\usepackage{graphics,graphicx,epsfig}
\usepackage{amssymb,color}
\usepackage{epsf,epstopdf,wrapfig}

\begin{document}

\date{today}

\newcommand{\tr}{{\mbox{\scriptsize {\sc T}}}}
\newcommand{\vek}[1]{\mathbf{#1}}

\title{Maximally informative ``stimulus energies'' in the analysis of\\ neural responses to natural signals}
\author{Kanaka Rajan and William Bialek}
\affiliation{Joseph Henry Laboratories of Physics and Lewis--Sigler Institute for Integrative Genomics, Princeton University, Princeton, New Jersey 08544}

\date{\today}

\begin{abstract}
The concept of feature selectivity in sensory signal processing can be formalized as dimensionality reduction: in a stimulus space of very high dimensions, neurons respond only to variations within some smaller, relevant subspace.  But if neural responses exhibit invariances, then the relevant subspace typically cannot be reached by a Euclidean projection of the original stimulus.  We argue that, in several cases, we can make progress by appealing to the simplest nonlinear construction, identifying the relevant variables as quadratic forms, or ``stimulus energies.''  Natural examples include non--phase--locked cells in the auditory system, complex cells in visual cortex, and motion--sensitive neurons in the visual system. Generalizing the idea of maximally informative dimensions, we show that one can search for the kernels of the relevant quadratic forms by maximizing the mutual information between the stimulus energy and the arrival times of action potentials.  Simple implementations of this idea successfully recover the underlying properties of model neurons even when the number of parameters in the kernel is comparable to the number of action potentials and stimuli are completely natural. We explore several generalizations that allow us to incorporate plausible structure into the kernel and thereby restrict the number of parameters.  We hope that this approach will add significantly to the set of tools available for the analysis of neural responses to complex, naturalistic stimuli.
 \end{abstract}
 
\maketitle

\section{Introduction}

A central concept in neuroscience is feature selectivity: as our senses are bombarded by complex, dynamic inputs, individual neurons respond to specific, identifiable components of these data \cite{barlow_72,barlow_95}.  Neurons early in a processing pathway are thought to be sensitive to simpler features \cite{barlow_53,lettvin_59}, and one can think of subsequent stages of processing as computing conjunctions of these features, so that neurons later in the pathway respond to more complex structures in the sensory world \cite{gross_02}.  A major challenge for theory is to make this intuition mathematically precise, and to use such a precise formulation to build tools that allow us to analyze real neurons as they respond to naturalistic inputs.  There is a long history of such work, but much of it rests on the identification of ``features'' with filters or templates.  Filtering is a linear operation, and matching to a template can be thought of as a cascade of linear and nonlinear steps.  As we will see, however, there are many examples of neural feature selectivity, well known from experiments on visual and auditory systems in many organisms, for which such a description in linear terms does not lead to much simplification.   

In this paper we use examples to motivate the simplest nonlinear definition of a feature, in which the relevant variable is a quadratic form in stimulus space.  Because the resulting variable is connected to the ``energy in frequency bands'' for auditory signals, we refer to these quadratic forms as ``stimulus energies.''  To be useful, we have to be able to identify these structures in experiments where neurons are driven by complex, naturalistic inputs.  We show that, generalizing the idea of maximally informative dimensions \cite{mid}, we can find the maximally informative stimulus energies using methods that don't require special assumptions about the structure of the input stimulus ensemble.  We illustrate these ideas on model neurons, and explore the amount of data that will be needed to use these methods in the analysis of real neurons.

\section{Motivation}

To motivate the problems that we address, let us start by thinking about an example from the auditory system.  This starting point is faithful to the history of our subject, since modern approaches for estimating receptive fields and filters have their origins in the classic work of de Boer and coworkers on the ``reverse correlation'' method \cite{deboer}, which was aimed at separating the filtering of acoustic signals by the inner ear from the nonlinearities of spike generation in  primary auditory neurons.  We will see that mathematically identical problems arise in thinking about complex cells in visual cortex,  motion sensitive neurons throughout the visual pathway, and presumably in other problems as well.

We begin with the simplest model of an auditory neuron. If the sound pressure as a function of time is $s(t)$, it is plausible that the activity of a neuron is controlled by some filtered version of this stimulus, so that the probability per unit time of generating a spike is
\begin{equation}
r(t) = r_0 g\left[ \int d\tau \, f(\tau ) s(t-\tau )\right] ,
\label{aud1}
\end{equation}
where $f(\tau )$ is the relevant temporal filter and $g[\cdot ]$ is a nonlinearity; the spikes occur at times   $t_{\rm i}$.   The statement that neurons are tuned is that if we look at the filter in Fourier space,
\begin{equation}
\tilde f(\omega ) = \int dt\,  f(t)  e^{i\omega t}  ,
\end{equation}
then the magnitude of the filter response, $| \tilde f(\omega )|$, has a relatively sharp peak near some characteristic frequency $\omega_c$.  If we choose the stimulus waveforms  from a Gaussian white noise ensemble, then the key result of reverse correlation is that if we compute the average stimulus in the neighborhood of a spike, we will recover the underlying filter, independent of the nonlinearity,
\begin{equation}
\langle s(t-t_{\rm i} ) \rangle \propto f(-t) .
\end{equation}
We emphasize that this is a theorem, not a heuristic data analysis method.  {\em If} the conditions of the theorem are met, then this analysis is guaranteed to give the right answer in the limit of large amounts of data.  If the conditions of the theorem are not met, then the spike--triggered average stimulus need not correspond to any particular characteristic of the neuron.

Eq.~(\ref{aud1}) is an example of dimensionality reduction. In principle, the neuron's response at time $t$ can be determined by the entire history of the stimulus for times $t' \leq t$.  Let us suppose that we sample (and generate) the stimulus in  discrete time steps spaced by $dt$.    Then the stimulus history is a list of numbers
\begin{equation}
{\vek s}_t \equiv \{ s(t), s(t-dt) , s(t- 2dt),  \cdots , s(t- Ddt)\},
\end{equation}
where $D$ is the effective stimulus dimensionality, set by $D=T/dt$, and $T$ the longest plausible estimate of the integration time for the neural response.  We can think of ${\vek s}_t$ as a $D$--dimensional vector.  If we know that the neural response is controlled by a linearly filtered version of the sound pressure stimulus, even followed by an arbitrary nonlinearity, then only one direction in this $D$--dimensional space matters for the neuron.   Further this really is a ``direction,''  since we can write the response as the Euclidean projection of $\vek s$ onto one axis, or equivalently the dot product between $\vek s$ and a vector $\vek W$,
\begin{equation}
r(t) = r_0 g({\vek W \cdot s}_t ),
\label{1D}
\end{equation}
where 
\begin{equation}
{\vek W} = dt \times \{ f(0) , f(dt) , f(2dt) , \cdots , f(T)\} .
\end{equation}
This explicit formulation in terms of dimensionality reduction suggests a natural generalization in which several dimensions, rather than just one, are relevant,
\begin{equation}
r(t) = r_0 g({\vek W_{\rm 1} \cdot s}_t, {\vek W_{\rm 2} \cdot s}_t , \cdots , {\vek W_{\rm K} \cdot s}_t ).
\label{Kdims}
\end{equation}
As long as we have $K\ll D$, it still holds true that the neuron responds only to some limited set of stimulus dimensions, but this number is not as small as in the simplest model of a single filter.

Notice that if an auditory neuron responds according to Eq (\ref{aud1}), then it will exhibit ``phase locking'' to periodic stimuli.  Specifically, if $s(t) = A\cos(\omega t)$ and $\tilde f (\omega ) =  | \tilde f (\omega )|e^{+i\phi}$, then $r(t) = r_0 g[ A | \tilde f (\omega )| \cos (\omega t - \phi )]$.   So long as there is a nonzero response to the stimulus, this response will be modulated at the stimulus frequency $\omega$, and more generally if we plot the spiking probability versus time measured by the phase $\psi = \omega t$ of the stimulus oscillation, then the probability will vary with, or ``lock'' to  this phase. 

While almost all auditory neurons are tuned, not all exhibit phase locking.  We often summarize the behavior of tuned, non--phase--locked neurons by saying that they respond to the power in a given bandwidth or to the envelope of the signal at the output of a filter.  The simplest model for such behavior, which has its roots in our understanding of hair cell responses \cite{javel, hudspeth, palmer, johnson}, is to imagine that the output of a linear filter passes through a weak nonlinearity, then another filter.  The second stage of filtering is low-pass, and will strongly attenuate any signals at or near the characteristic frequency $\omega_c$.  Then, to lowest order, the neuron's response depends on
\begin{equation}
p(t) = \int d\tau\, f_2 (\tau ) \left[ \int dt' f_1 (t-\tau - t') s(t')\right]^2 ,
\label{aud_power}
\end{equation}
where $f_1$ is the bandpass filter that determines the tuning of the neuron and $f_2$ is a smoothing filter which ensures that the cell responds to the power in its preferred frequency band while rejecting any temporal variation at or above the ``carrier'' frequency.  The probability of spiking depends on this power $p(t)$ through a nonlinearity, as before,
\begin{equation}
r(t) = r_0\, g[p(t)].
\label{spk_prob}
\end{equation}
\begin{figure*}[!ht]
\begin{center}
\includegraphics[width=5.5in]{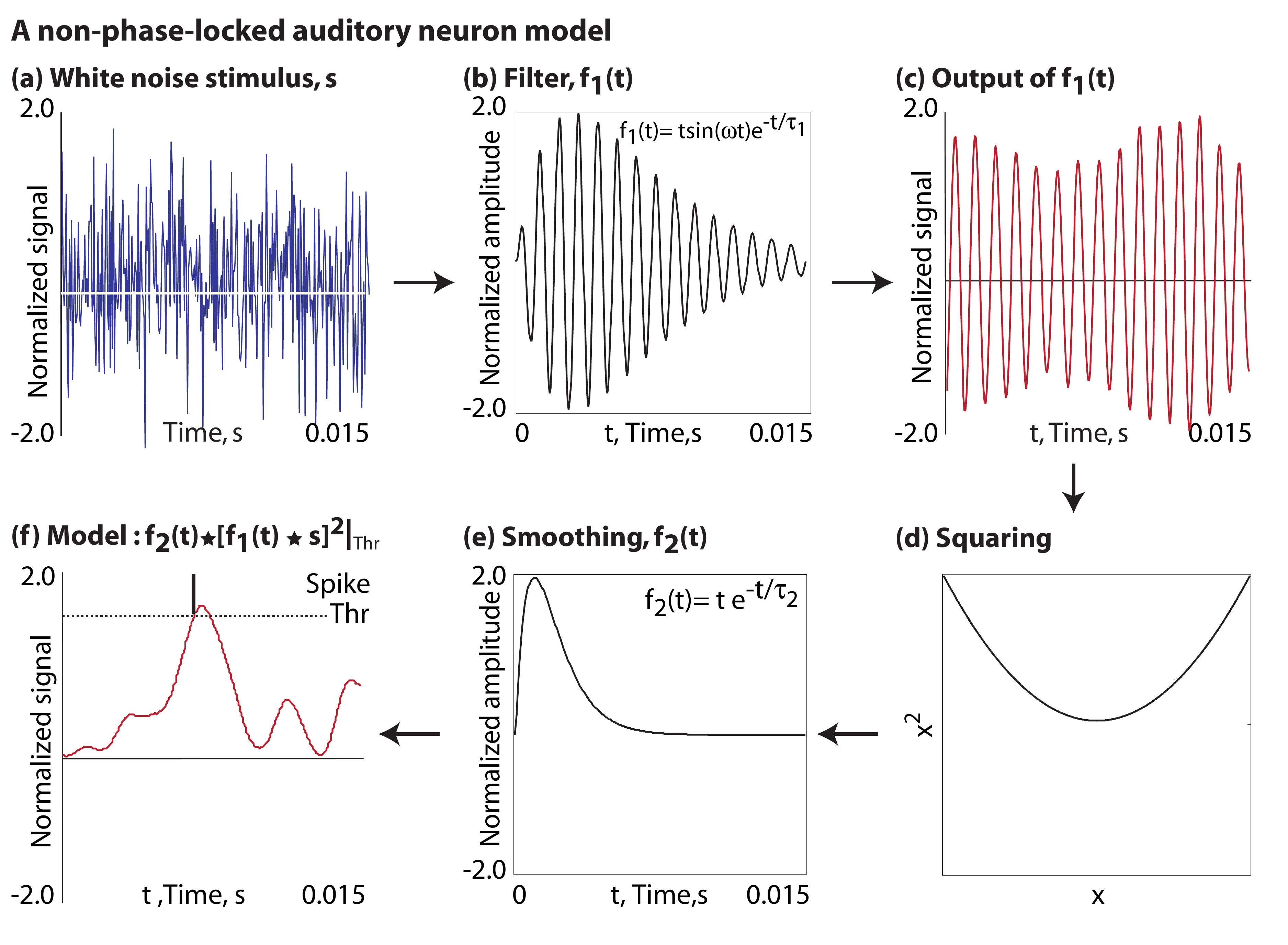}
\end{center}
\caption{{\bf A non--phase--locked auditory neuron. (a)} In this implementation a model neuron responds to a white noise stimulus.   {\bf(b)} The stimulus $\vek s$ is filtered through a temporal filter $f_1$, which has the form $t\sin(\omega t)\exp(-t/\tau_1)$ where $\omega = 2\pi\times 10^{3} \,{\rm Hz}$ and $\tau_1 = 3\, {\rm ms}$. {\bf(c)} The output of $f_1$ is shown here. The filter $f_1$ is narrow band, therefore the output oscillates at the characteristic frequency even when the input is white. {\bf(d)} The output of $f_1$ is first squared, and in {\bf(e)}, convolved with a second filter $f_2$ of the form $t\exp(-t/\tau_2)$, with a smoothing time constant $\tau_2 = 1$ ms. {\bf(f)} The normalized signal is finally thresholded to generate spikes at a mean rate that allows the spiking probability to be $0.1$ per time step. We assume that time runs discretely in steps of $dt = 1/20000\, {\rm s}$.}
\label{f1}
\end{figure*}
\begin{figure*}[!ht]
\begin{center}
\includegraphics[width=5.75in]{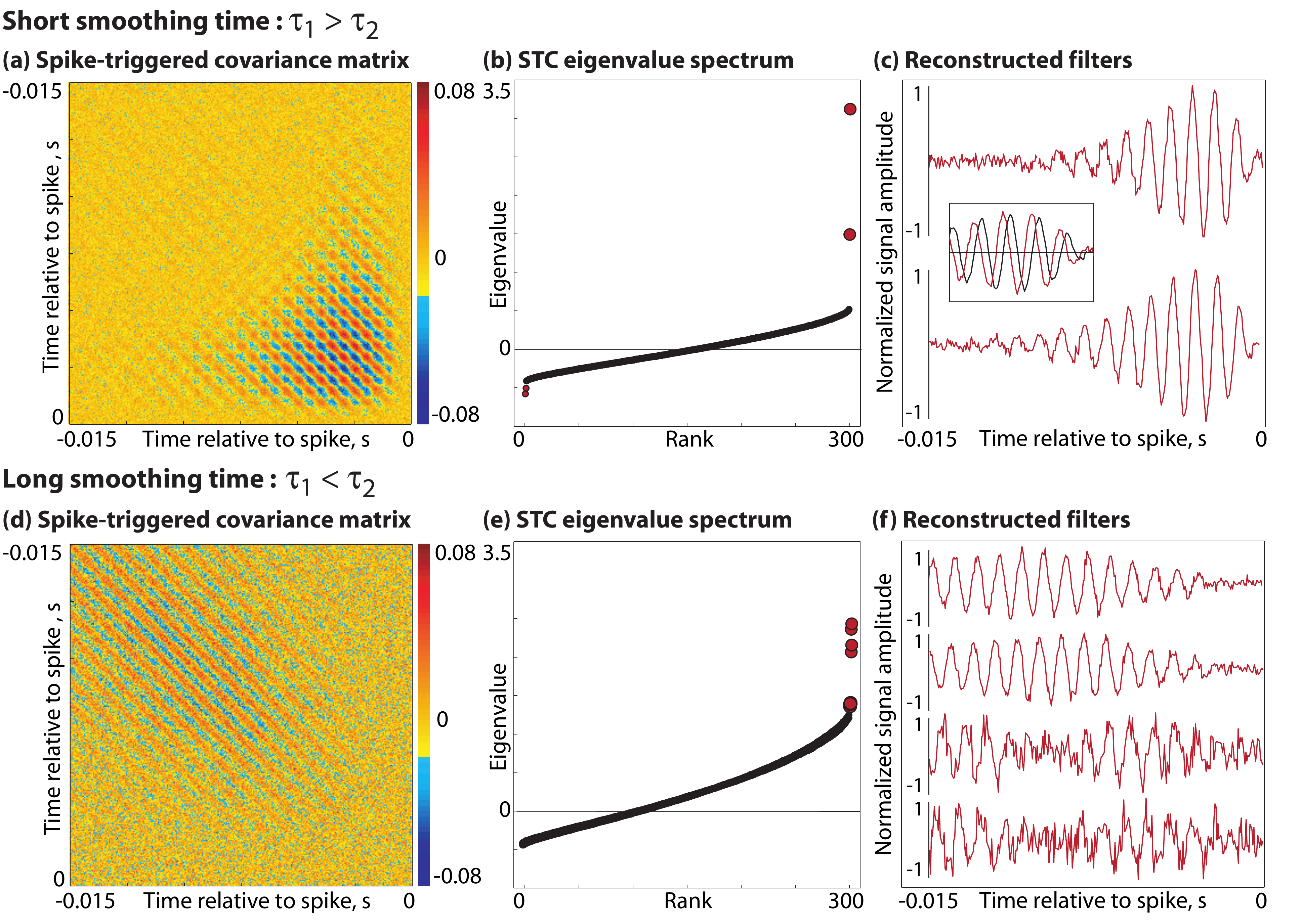}
\end{center}
\caption{{\bf Covariance analysis of the non--phase--locked auditory neuron.} {\bf(a)} If the time constant of the smoothing filter $f_2$ is much shorter than that of the filter $f_1$, the spike--triggered covariance matrix has a relatively simple structure. {\bf(b)} Diagonalizing this covariance matrix yields $2$ leading eigenvalues (the rest remain close to $0$). {\bf(c)} The eigenvectors corresponding to the $2$ non-zero eigenvalues are the reconstructed filters plotted here. These form a quadrature pair, as shown in the inset. {\bf(d)} If the smoothing time of the filter $f_2$ is larger than that of the first, the covariance matrix has a much richer structure. {\bf(e)} The spike--triggered covariance matrix decomposes into multiple non-unique eigenvalues. {\bf(f)} The eigenvectors corresponding to the non-zero eigenvalues give multiple time-shifted copies of the same filter.}
\label{f2}
\end{figure*}

Intuitively, this simple model for a non--phase--locked neuron also represents a substantial reduction in dimensionality -- all that matters is the power passing through a given frequency band, defined by the filter $f_1$.  On the other hand, we cannot collapse this model into the one dimensional form of Eq.~(\ref{1D}).   To be concrete, suppose that the filter $f_1$ has a relatively narrow bandwidth around its characteristic frequency $\omega_c$.  Then we can write
\begin{equation}
f_1(\tau ) = A(\tau ) \sin (\omega_c \tau+ \phi) ,
\end{equation}
where the amplitude $A(\tau )$ varies slowly compared to the period of the oscillation.  Let us denote by $\tau_1$ the temporal width of $A(\tau)$, since this corresponds to the time over which the filter $f_1$ integrates the stimulus, and similarly $\tau_2$ will denote the temporal width of $f_2(\tau)$. To make sure that the power $p(t)$ does not oscillate at (twice) the frequency $\omega_c$, we must force $\tau_2 \gg 2\pi/\omega_c$.  But we still have two possibilities, (a) $\tau_1 \gg \tau_2 \gg \omega_c\;$ and (b) $\tau_2 \gg \tau_1 \gg \omega_c$. If (a) is true, we can show that $p(t)$ is the Pythagorean sum of the outputs of two filters that form a quadrature pair,
\begin{eqnarray}
p(t) &\approx&  \left[ \int dt' f_\alpha  (t-\tau - t') s(t')\right]^2  \nonumber\\
&&\,\,\,\,\,\,\,\,\,\, + \left[ \int dt' f_\beta  (t-\tau - t') s(t')\right]^2 ,
\end{eqnarray}
where
\begin{eqnarray}  
f_\alpha (\tau) &\approx&   A(\tau) \sin(\omega_c \tau + \phi) \\
f_\beta (\tau) &\approx&   A(\tau) \cos(\omega_c \tau + \phi) .
\end{eqnarray}
On the other hand, if (b) is true, there is no simple decomposition, and   the minimum number of dimensions that we need to describe this model is $K\sim \tau_2/dt$, which can be quite large.  

We can measure the number of relevant dimensions using the spike--triggered covariance matrix.    Specifically, if the stimulus vector ${\vek s}_t$ has components $s_{\rm i} (t)$, then we can form the matrix 
\begin{equation}
\Delta C_{\rm ij} = \langle s_{\rm i}(t) s_{\rm j}(t)\rangle_{t = t_{\rm spike}} -  \langle s_{\rm i}(t) s_{\rm j}(t)\rangle ,
\end{equation}
where in the first term we average over the arrival time of the spikes and in the second term we average over all time.  If the spiking probability behaves as in Eq.~(\ref{Kdims}), and we choose the stimulus from a Gaussian ensemble, then $\Delta C$ has exactly $K$ nonzero eigenvalues \cite{fnd, blaise+adrienne}.  In Fig.~\ref{f1} we schematize the model auditory neuron we have been describing, and in Fig.~\ref{f2} we show the spike--triggered covariance analysis for models in the two limits, $\tau_2 \ll \tau_1$ and $\tau_2 \gg \tau_1$.  Indeed we find that in the first case there are just two relevant dimensions, a quadrature pair of filters, whereas in the second case there are many relevant dimensions; these dimensions appear as temporally shifted and orthogonalized copies of the filter $f_1(t)$. 

We can think of a neuron that does not phase lock as having an invariance:  it responds to acoustic waveforms that have energy in a relevant bandwidth near $\omega_c$,  but it doesn't discriminate among signals that are shifted by small times.  This invariance means that the cell is not just sensitive to one dimension of the stimulus, but to many, although these different dimensions correspond, in effect, to the same stimulus feature occurring at different times relative to the spike.  Thus, we have a conflict between the notion of  a ``single feature'' and the mathematical description of  a ``single dimension'' via linear projection.  The challenge is to provide a mathematical formulation that better captures our intuition. Before presenting a possible solution, let's see how the same problem arises in other cases.

Since the classical work of Hubel and Wiesel \cite{hubelwiesel1, hubelwiesel2}, we know that cells in the primary visual cortex can be classified as ``simple'' and ``complex.''   Although Hubel and Wiesel did not give a mathematical description of their data, in subsequent work, simple cells often have been described in the same way that we described the simplest auditory neuron in Eq (\ref{aud1}) \cite{recio}.  If the light intensity falling on the retina varies in space ($\vek{\vec x}$) and time ($t$) as $I({\vek{ \vec x}}, t)$, we can define a spatiotemporal receptive field $F({\bf \vec x}, \tau )$ and approximate the probability that a simple cell generates a spike per unit time as
\begin{equation}
r(t) = r_0 \;g\left[ \int d^2 x \int d\tau \;F({\bf \vec x}, \tau )\; I({\bf \vec x}, t) \right].
\label{simple1}
\end{equation}
If, as before, we assume that the stimulus is generated in discrete time steps (movie frames) with spacing $dt$, and that the stimulus influences spikes only within some time window of duration $T$, then we can think of the stimulus at any moment in time as being the $T/dt$ frames of the movie preceding that moment,
\begin{equation}
\!{\vek s}_t\! \equiv\! \{ I({\bf \vec x}, t), I({\bf \vec x}, t-dt) , I({\bf \vec x}, t- 2dt),\cdots, I({\bf \vec x}, t-T)\}. 
\end{equation}
If the relevant region of space is within $d\times d$ pixels, then this stimulus vector lives in a space of dimension $D = d^2 T/(dt)$, which can be enormous.  As in the discussion above, Eq.~(\ref{simple1}) is a restatement of the hypothesis that only one direction in this space is relevant for determining the probability that the simple cell generates a spike, and ``direction'' is once again a Euclidean or linear projection.  

For a complex cell, on the other hand, this single projection is inadequate. Complex cells respond primarily to oriented edges and gratings, as do simple cells, but they have a degree of spatial invariance which means that their receptive fields cannot be mapped onto fixed zones of excitation and inhibition. Instead, they respond to patterns of light in a certain orientation within a large receptive field, regardless of precise location, or to movement in a certain direction.   Corresponding  to this intuition, analysis of complex cells using the spike--triggered covariance method shows that there is more than one relevant dimension \cite{rustV1}.  As with non--phase--locked auditory neurons, what defeats the simplest version of dimensionality reduction in complex cells is the invariance of the response, in this case, invariance to small spatial displacement of the relevant, oriented stimulus feature.

The simplest model of a complex cell is precisely analogous to the quadrature pair of filters that emerge in the analysis of non--phase--locked auditory neurons.  To be concrete, let us imagine that receptive fields are described by Gabor patches.  The position $\bf \vec x$ includes two orthogonal coordinates in visual space, which we call $x_1$ and $x_2$.  Gabor patches have an oscillatory dependence on one of these coordinates, but simply integrate along the other; both the integration and the envelope of the oscillations are described by Gaussians, so that
\begin{equation}
F_1 ({\bf \vec x}, \tau ) = \cos(k x_1) \exp\left[ - \left( {{x_1^2}\over{2\sigma_1^2}} +  {{x_2^2}\over{2\sigma_2^2}}\right)\right] f(\tau) ,
\end{equation}
and the quadrature filter is then
\begin{equation}
F_2 ({\bf \vec x}, \tau ) = \sin(k x_1) \exp\left[ - \left( {{x_1^2}\over{2\sigma_1^2}} +  {{x_2^2}\over{2\sigma_2^2}}\right)\right]f(\tau) .
\end{equation}
Each of these filters is maximally sensitive to extended features oriented along the $x_2$ direction, but the optimal patterns of light and dark are shifted for the two filters; for simplicity we have assumed that spatial and temporal filtering is separable.  If we form the energy--like quantity
\begin{eqnarray}
p(t) &=& \left[ \int d^2 x \int d\tau \;F_1({\bf \vec x}, \tau )\; I({\bf \vec x}, t) \right]^2\nonumber\\
&&\,\,\,\,\,\,\,\,\,\, + \left[ \int d^2 x \int d\tau \;F_2({\bf \vec x}, \tau )\; I({\bf \vec x}, t) \right]^2,
\label{complexcell_power}
\end{eqnarray}
we have a measure of response to oriented features independent of their precise position, and this provides a starting point for the analysis of a complex cell.

One more example of the problem we face is provided by the computation of motion in the visual system.  There is a classical model for this computation, the correlator model, that grew out of the experiments by Hassenstein and Reichardt on behavioral responses to visual motion in beetles and flies \cite{hassenstein}.  Briefly, the idea behind this model is that if something is moving at velocity $v$, then the image intensity $I(x,t)$ must be correlated with the intensity at $I(x+\Delta , t +\tau)$, where $\tau = \Delta/v$.  Then we can detect motion by computing this correlation and averaging over some window in space and time,
\begin{widetext}
\begin{equation}
{\vek C}_{\Delta,\tau} (t) = \int dx \, W(x) \int dt' f(t-t') I(x+\Delta , t' ) I(x+\Delta , t'+\tau ) ,
\end{equation}
where for simplicity, we think just about a single spatial dimension.  In principle, with just one value of the delay $\tau$ and one value of the displacement $\Delta$, this correlation ``detects'' only motion at one velocity, $v = \Delta/\tau$, but we can easily generalize this computation to a weighted sum over different values of these spatiotemporal shifts, 
\begin{equation}
{\vek C} (t) = \int dx \, W(x) \int dt' f(t-t') \int d\tau  \sum_{\Delta} F(\Delta , \tau ) I(x+\Delta , t' ) I(x+\Delta , t'+\tau ).
\label{reichardt2}
\end{equation}
Depending on the precise form of this weighting we can arrange for the correlation $\vek C$ to have a relatively smooth, graded dependence on the velocity.  
\end{widetext}

In the insect visual system it seems natural to think of the correlations in Eq.~(\ref{reichardt2}) as being computed from the outputs of individual photoreceptors, which are typically spaced $\sim 1^\circ$ apart.  In mammals, we can imagine computing a similar quantity, but we would use the outputs of larger retinal or cortical receptive fields \cite{adelson}.  We can also think about this computation in Fourier space.  If we transform 
\begin{equation}
I(x,t) = \int {{dk}\over{2\pi}}\int {{d\omega}\over{2\pi}}\; \tilde I (k,\omega)\; e^{ikx - i \omega t},
\end{equation}
then we can think of $\vek C$ as integrating power or energy $P(k,\omega ) = | \tilde I (k,\omega)|^2$ over some region in  the $k-\omega$ plane;  motion corresponds to having this power concentrated along the line $\omega = vk$.  Once again there is an invariance to this computation, since as with the non--phase--locked auditory neuron we are looking for power regardless of phase.  More directly, in this case the brain is trying to compute a velocity, more or less independently of the absolute position of the moving objects.  Even in an insect visual system, where computing $\vek C$ corresponds to correlating the filtered outputs of neighboring photoreceptors in the compound eye, this computation is repeated across an area containing many photoreceptors, and hence there is no way to collapse  $\vek C$ down to a function of just one or two Euclidean projections in the stimulus space. 

What do these examples -- non--phase--locked auditory neurons, complex cells in the visual cortex, and visual motion detectors -- have in common?  In all three cases, the natural, simplest starting point is a model in which the brain computes not a linear projection of the stimulus onto a receptive field, but rather a quadratic form.  More precisely, if stimuli are the vectors ${\vek s}\equiv \{s_1, s_2, \cdots s_D\}$, then Eq's.~(\ref{aud_power}),~(\ref{complexcell_power}), and ~(\ref{reichardt2}) all correspond to computing a quantity
\begin{equation}
x = {\vek s}^\tr \cdot \vek Q \cdot \vek s \equiv \sum_{{\rm i,j} = 1}^D s_{\rm i}\; Q_{\rm ij}\; s_{\rm j} .
\label{stim_eng1}
\end{equation}
Generalizing the use of terms such as ``energy in a frequency band'' and ``motion energy,'' we refer to $x$ as a ``stimulus energy.''

If the matrix $\vek Q$ is of low rank, this means that we can build $x$ out of projections of the stimulus onto a correspondingly low dimensional Euclidean subspace, and we can try to recover the relevant structure using methods such as the spike--triggered covariance or maximally informative dimensions.  Still, if $\vek Q$ is of rank $5$ (for example), once we identify the five relevant dimensions we have to explore the nonlinear input/output relation of the neuron thoroughly enough to recognize whether these dimensions combine in a simple way, to recover our intuition that there is a single stimulus feature $x$ to which the cell responds. This can be prohibitively difficult.  

In many cases, it is plausible that neurons could be responding just to one energy $x$, but the underlying matrix $\vek Q$ could be of full rank; a clear example is provided by correlator models for wide--field motion sensitive neurons, as in \cite{hassenstein, fnd}.  But in this case there is no real ``dimensionality reduction'' associated with the mapping from $\vek s \rightarrow x$, if all we know how to do is to search for linear projections or Euclidean subspaces.  On the other hand, mapping ${\vek s} \rightarrow x$ really is a tremendous reduction in complexity, because the full stimulus $\vek s$ is described by $D$ parameters, while $x$ is just one number.  

Suppose that the response of a neuron to complex stimuli can be described by saying that the probability of spiking depends solely on a single stimulus energy $x$ as in Eq.~(\ref{stim_eng1}), so that
\begin{equation}
r(t) = r_0 g( {\vek s}_t^\tr \cdot \vek Q \cdot {\vek s}_t) .
\label{r_sQs}
\end{equation}
Our task becomes one of showing that we can recover the underlying matrix $\vek Q$ by analyzing the spike train in relation to the stimuli, without making any assumptions about the statistics of the stimuli.  

As we were finishing this work, we became aware of two other recent efforts that point to the importance of quadratic forms in stimulus space.  It is shown in \cite{fitzgerald+al_11} that a logistic dependence of the spike probability on a quadratic form is the maximum entropy, and hence least structured, model consistent with a measurement of the spike--triggered covariance matrix, and that this identification is correct without any assumptions about the distributions of inputs.  As a practical matter, Ref \cite{fitzgerald+al_11} emphasizes the case where the matrix kernel (corresponding to $\bf Q$ above) is of low rank, and explores examples in which the quadratic form provides an incremental improvement on linear dimensionality reduction.  In a different direction, the work in \cite{park+pillow_11} consider models in which the spike probability depends exponentially on a quadratic form, and spiking is explicitly a Poisson process.  They show that this model, and some generalizations, lends itself to a Bayesian formulation, in which various simplifying structures (see Section IV) can be imposed through prior distributions on the possible $\bf Q$, although it is not clear whether computationally tractable priors correspond to realistic models.  In some limits, these approaches are equivalent to one another, and to the search for maximally informative stimulus energies that we propose here.  As far as we can see, the Maximally Informative Stimulus Energy method always involves fewer assumptions, and can capture any of the structures postulated in the other methods.

\section{Core of the method}

If the probability of generating an action potential depends on the stimulus $\vek s$, then observing the arrival of even a single spike provides information about the stimulus.  Importantly, the data processing inequality \cite{cover+thomas_91} tells us that if we look not at the full stimulus but only at some limited or compressed description of the stimulus -- a single feature, for example -- we can only lose information.  If the neuron really is sensitive to only one stimulus feature, however, it is possible to lose none of the available mutual information between spikes and stimuli by focusing on this one feature.  This suggests a natural strategy for identifying relevant stimulus features, by searching for those which preserve the maximum amount of information \cite{mid}.  Further, we can put the success of such a search on an absolute scale, by estimating more directly the information that spikes provide about the stimulus \cite{strong+al_98, brenner+al_00}, and asking what fraction of this information is captured by the best feature we could find.  

\subsection{Setting up the problem}

To make this precise, we recall that the information about the stimulus $\vek s$ that is conveyed by the observation of a single spike can be written, following \cite{brenner+al_00}, as 
\begin{equation}
I_{\rm spike}   = \int d^D {\vek s}\; P({\vek s}|{\rm spike}) \log_2\left[ {{P({\vek s}|{\rm spike})}\over{P({\vek s})}}\right] \,{\rm bits},
\end{equation}
where $P({\vek s})$ is the stimulus distribution and $P({\vek s}|{\rm spike})$ is the distribution of stimuli given that a spike occurred at a particular time (the response conditional ensemble \cite{stc}).  If we consider a mapping ${\cal M}: {\vek s} \rightarrow x$, then we can also compute
\begin{equation}
I_{\rm spike}  ({\cal M}) = \int dx \;P(x|{\rm spike}) \log_2\left[ {{P(x|{\rm spike})}\over{P(x)}}\right] , 
\end{equation}
 knowing that for any mapping $\cal M$, $I_{\rm spike}({\cal M}) \leq I_{\rm spike}$.    If we restrict ourselves to some class of mappings, then we can search for an ${\cal M}^*$ which maximizes $I_{\rm spike}({\cal M^*})$, and see how close this is to the real value of $I_{\rm spike}$.   If the inputs are Gaussian, then we can find $\cal M$ by standard spike--triggered or reverse correlation methods. Working with arbitrary (i.e., natural) stimulus ensembles complicates things. Further, if there is just one stimulus feature $x$, then computing $I_{\rm spike}({\cal M})$ involves only probability distributions over a single variable, so there is no curse of dimensionality.  Finally, in order for this to work, we need to make no assumptions about the form of the distribution of stimuli $P({\vek s})$, or the inherited distribution of stimulus features $P(x)$.  Thus, as first emphasized in \cite{mid}, searching for maximally informative features provides a practical and principled way to analyze neural responses to high dimensional, naturalistic stimuli.

The work in \cite{mid} considered the case where the stimulus features are one or more  linear projections of the stimulus, $x_\alpha = {\vek W}_{\alpha} { \cdot \vek s}$, so that the mapping ${\cal M}$ is parameterized by the vectors $\{{\vek W}_\alpha\}$; in the simplest case there being just one vector $\vek W$.    Here we are interested in quadratic mappings, corresponding to the stimulus energies in Eq.~(\ref{stim_eng1}).  Now the mapping $\cal M$ is parameterized by a symmetric matrix $\vek Q$.  In principle, all the arguments of \cite{mid} for the linear case should generalize to the quadratic case, and we follow this path.  Before starting, we note an obvious problem related to the number of parameters we are looking for.  If we are searching for a vector $\vek W$ in a $D$--dimensional stimulus space, we are looking an object described by $D$ numbers.  In fact, the length of the vector is irrelevant, so that maximizing $I_{\rm spike}  ({\cal M})$ corresponds to optimization in a $D-1$ dimensional space.  But if we are searching for a symmetric matrix that acts on stimulus vectors, there are $D(D+1)/2 - 1$ free parameters.  This is a problem both because we have to optimize in a space of much higher dimensionality, and because determining more parameters reliably must require larger data sets.  We will address these problems shortly, but let's start by following the path laid out in \cite{mid}, which involves searching for the maximum of $I_{\rm spike}  ({\cal M})$ by gradient ascent.

If our stimulus feature is the energy in Eq.~(\ref{stim_eng1}), then the distribution of $x$ is
\begin{equation}
P_{\vek Q}(x) = \int d^D s \; \delta (x- {\vek s}^\tr \cdot \vek Q\cdot \vek s) \;P({\vek s}) ,
\end{equation}
where the subscript explicitly denotes that $P(x)$ depends on $\vek Q$.  We take the derivative of this with respect to an element $Q_{\rm ij}$ in the matrix $\vek Q$,
\begin{eqnarray}
{{\partial P_{\vek Q}(x)}\over{\partial Q_{\rm ij}}} &=& \int d^D s \; {\partial\over{\partial Q_{\rm ij}}} \delta (x- {\vek s}^\tr  \cdot \vek Q\cdot \vek s) P({\vek s})\\
&=&
- \int d^D s \;  s_{\rm i} s_{\rm j}\; \delta ' (x- {\vek s}^\tr \cdot\vek Q \cdot \vek s) \;P({\vek s})  .
\end{eqnarray}
Similarly, we can differentiate the distribution of $x$ conditional on a spike,
\begin{equation}
{{\partial P_{\vek Q}(x|{\rm spike})}\over{\partial Q_{\rm ij}}}  = - \int d^D s \; s_{\rm i} s_{\rm j}\; \delta ' (x- {\vek s}^\tr  \cdot \vek Q\cdot \vek s)\; P({\vek s}|{\rm spike}) .
\end{equation}
Putting these terms together, we can differentiate the information:
\begin{widetext}
\begin{eqnarray}
{{\partial I_{\rm spike}  ({\cal M}) }\over {\partial Q_{\rm ij}}} &=& \int dx \left[ {{\delta I_{\rm spike}  ({\cal M})}\over{\delta P_{\vek Q}(x)}} {{\partial P_{\vek Q}(x)}\over{\partial Q_{\rm ij}}} + {{\delta I_{\rm spike}  ({\cal M})}\over{\delta P_{\vek Q}(x|{\rm spike})}} {{\partial P_{\vek Q}(x|{\rm spike})}\over{\partial Q_{\rm ij}}}\right]\\
&=& {1\over{\ln 2}} \int dx \, {{P_{\vek Q}(x|{\rm spike})}\over{P_{\vek Q}(x)}}\int d^D s \, s_{\rm i} s_{\rm j} \delta ' (x- {\vek s}^\tr { \cdot \vek Q\cdot \vek s}) P({\vek s})\nonumber\\
&&\,\,\,\,\,\,\,\,\,\, - {1\over{\ln 2}} \int dx \left( 1 + \ln \left[ {{P_{\vek Q} (x|{\rm spike})}\over{P_{\vek Q} (x)}}\right]\right) \int d^D s \, s_{\rm i} s_{\rm j} \delta ' (x- {\vek s}^\tr { \cdot \vek Q\cdot \vek s}) P({\vek s}|{\rm spike}) \\
&=& - {1\over{\ln 2}} \int dx \int d^D s \, s_{\rm i} s_{\rm j} \delta (x- {\vek s}^\tr { \cdot \vek Q\cdot s}) P({\vek s})
{d\over{dx}}\left[  {{P_{\vek Q}(x|{\rm spike})}\over{P_{\vek Q}(x)}}\right]\nonumber\\
&&\,\,\,\,\,\,\,\,\,\,  + {1\over{\ln 2}} \int dx {{P_{\vek Q}(x)}\over{P_{\vek Q}(x|{\rm spike})}}  \int d^D s \, s_{\rm i} s_{\rm j} \delta (x- {\vek s}^\tr { \cdot \vek Q\cdot \vek s}) P({\vek s}|{\rm spike}) {d\over{dx}}\left[  {{P_{\vek Q}(x|{\rm spike})}\over{P_{\vek Q}(x)}}\right] .
\end{eqnarray}
But we notice that 
\begin{equation}
 \int d^D s \; s_{\rm i} s_{\rm j}\; \delta (x- {\vek s}^{\tr} { \cdot \vek Q\cdot \vek s}) P({\vek s}) = \langle s_{\rm i} s_{\rm j} | x\rangle \;P_{\vek Q}(x) ,
\end{equation}
where $\langle s_{\rm i} s_{\rm j} | x\rangle$ is the expectation value of $s_{\rm i}s_{\rm j}$ conditional on the value of the stimulus energy $x$, and similarly
\begin{equation}
 \int d^D s \, s_{\rm i} s_{\rm j} \delta (x- {\vek s}^{\tr} { \cdot \vek Q\cdot \vek s}) P({\vek s}|{\rm spike}) = \langle s_{\rm i} s_{\rm j} | x,{\rm spike}\rangle P_{\vek Q}(x|{\rm spike}) ,
\end{equation}
where $\langle s_{\rm i} s_{\rm j} | x, {\rm spike}\rangle$ is the expectation value conditional on the energy $x$ {\em and} the occurrence of a spike.  We can combine these terms to give
\begin{equation}
{{\partial I_{\rm spike}  ({\cal M}) }\over {\partial Q_{\rm ij}}}  = \int dx\, P_{\vek Q}(x) \left[ \langle s_{\rm i} s_{\rm j} | x, {\rm spike}\rangle - \langle s_{\rm i} s_{\rm j} | x\rangle\right]  {d\over{dx}}\left[  {{P_{\vek Q}(x|{\rm spike})}\over{P_{\vek Q}(x)}}\right] ,
\end{equation}
or, more compactly,
\begin{equation}
\nabla_{\vek Q}  I    = \int dx\, P_{\vek Q}(x) \left[ \langle {\vek s \vek s}^\tr  | x, {\rm spike}\rangle - \langle {\vek s \vek s}^\tr  | x\rangle\right]  {d\over{dx}}\left[  {{P_{\vek Q}(x|{\rm spike})}\over{P_{\vek Q}(x)}}\right] .
\label{IGrad}
\end{equation}
To learn the maximally informative energy, or the best choice  of the matrix $\vek{Q}$, we can ascend the gradient in successive learning steps,
\begin{equation}
\vek Q \rightarrow \vek Q+\gamma\;\nabla_{\vek Q} I \mathrm{\;\;where\; \gamma \;is \;small.}
\label{grad_ascent}
\end{equation}
\end{widetext}

\subsection{Multiple matrices}

In the same way that the idea of linear projection can be generalized to have the probability of spiking depend on multiple linear projections,  we can generalize to the case where the are multiple relevant stimulus energies.  Perhaps the simplest example Eq.~(\ref{r_sQs}) can be generalized to $2$ matrices, is the computation of a (regularized) ratio between two stimulus energies, so that the probability of spiking varies as
\begin{equation}
r  = r_0 g\left(\frac{\vek{s^\tr}\cdot \vek Q_1\cdot\vek s}{1 + \vek {s^\tr}\cdot \vek Q_2\cdot\vek s}\right).
\label{dual}
\end{equation}
Some biological examples of this formulation include gain control or normalization in V1 \cite{rustV1}, optimal estimation theory of motion detection in visual neurons of insects \cite{fnd} and complex spectrotemporal receptive fields of neurons responsible for song processing in songbirds \cite{sahani, theuni}. The inference task becomes one of estimating both matrices $\vek Q_1$ and $\vek Q_2$ by information maximization. 

\begin{widetext}
As before, we can compute the gradient, noting that this time, there are two different gradients of $I_{\rm spike}(\cal M)$,
\begin{eqnarray}
\!\nabla_{\vek Q_1}I &=& \int\!\! dx\;P_{\vek{Q_1}}(x)\left[\left\langle\left.\frac{\vek s \vek s^\tr}{1 + \vek s^\tr \cdot\vek Q_2\cdot\vek s} \right|x, \mathrm{spike}\right\rangle-\left\langle\left.\frac{\vek s \vek s^\tr}{1 + \vek s^\tr\cdot \vek Q_2\cdot\vek s} \right|x\right\rangle\right]\frac{d}{dx}\left[\frac{P_{\vek Q_1}(x|\mathrm{spike})}{P_{\vek Q_1}(x)}\right]\;\mathrm{and\;},\nonumber\\
\!\nabla_{\vek Q_2}I &=& -\int\!\! dx\;P_{\vek{Q_2}}(x)\left[\left\langle\left.\frac{\vek s^\tr \cdot\vek Q_1\cdot\vek s}{\left(1 + \vek s^\tr\cdot \vek Q_2\cdot\vek s\right)^2}\;\vek s \vek s^\tr \right|x, \mathrm{spike}\right\rangle-\left\langle\left.\frac{ \vek s^\tr \cdot \vek Q_1\,\cdot\vek s}{\left(1 + \vek s^\tr\cdot \vek Q_2\cdot\vek s\right)^2}\;\vek s \vek s^\tr \right|x\right\rangle\right]\frac{d}{dx}\left[\frac{P_{\vek Q_2}(x|\mathrm{spike})}{P_{\vek Q_2}(x)}\right].
\label{dualGrad}
\end{eqnarray}
Analogous to Eq.~(\ref{grad_ascent}), at every learning step, we update each matrix $\vek Q_i$ by the appropriate $i^{\mathrm {th}}$ gradient,
\begin{equation}
\vek Q_i \rightarrow \vek Q_i + \gamma\,\nabla_{\vek Q_i}I.\label{dualGrad2}
\end{equation}
where $i = 1, 2$ for the $x$ in Eq.~(\ref{dual}). In principle the formalism in Eq.~(\ref{dual}) could yield a more complete description of a neuron's nonlinear response properties compared to a single quadratic kernel convolving the stimulus, but there are some data-requirement challenges which we will address later.
\end{widetext}

\subsection{Technical aspects of optimization}

In order to implement Eq.~(\ref{grad_ascent}) as an algorithm, we have to evaluate all the relevant probability distributions and integrals.  In practice, this means  computing $x$ for all stimuli, choosing an appropriate binning along the $x$--axis, and sampling the binned versions of the spike--triggered and prior distributions. We compute the expectation values $\langle \vek{s}\vek s^\tr\rangle $ separately for each bin, approximate the integrals as sums over the bins, and derivatives as differences between neighboring bins. To deal with local extrema in the objective function, we use a  large starting value of $\gamma$ and gradually decrease $\gamma$ during learning. This basic prescription can be made more sophisticated, but we do not report these technical improvements here. 
 \begin{figure*}[bht]
\begin{center}
\includegraphics[width=6.25in]{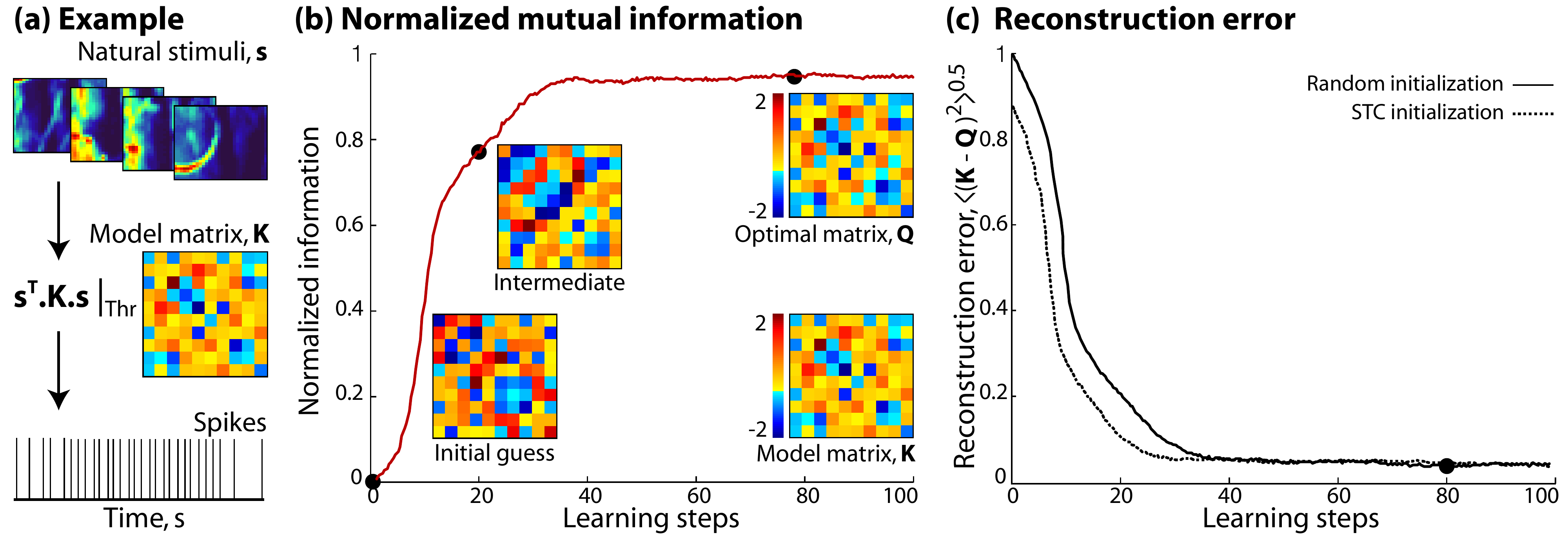}
\end{center}
\caption{
{\bf Core of the method.} {\bf(a)} A general implementation is shown here. The stimuli $\vek s$ are natural image clips which are $D\times D$ pixel patches resized from a natural image database, as described in \cite{pidb}. $1000$ spikes are generated with a probability per time bin of $0.1$ from the model neuron by a thresholding the term, $\vek{s^\tr \cdot K\cdot s}$ where the $10\times 10$ matrix $\vek K$ is the receptive field of the neuron. {\bf(b)} Mutual information between the spiking response of the model neuron and the quadratic stimulus projection $x$ is plotted as a function of the number of learning steps. Information, normalized by its value when $\vek K = \vek Q$, peaks at the $40^{\mathrm{th}}$ learning step and then plateaus. The $3$ black dots on the trace denote the points at which we extract the initial, the intermediate and the optimal matrices. The maximally informative matrix $\vek Q$ reconstructed at the $40^{\mathrm{th}}$ step, agrees well with $\vek K$, indicating convergence. For this implementation the step size $\gamma = 0.5$ at the start and $0.05$ at the end of the algorithm. {\bf(c)} Root--mean--square (RMS) reconstruction error calculated as $\langle(\vek K - \vek Q)^2\rangle^{1/2}$, is plotted as a function of the number of learning steps. This error decreases steadily until either the randomly initialized matrix (solid line) or the matrix initialized to the spike--triggered covariance matrix (dashed line) matches $\vek K$. If $\vek Q$ is initialized to the covariance matrix, the initial RMS error is smaller and the convergence is faster ($30^{\mathrm{th}}$ learning step) compared to that for a randomly initialized $\vek Q$. For this example, both $\vek K$ and $\vek Q$ are $10\times 10$ matrices and the black dot on the solid trace is at the same learning step as in panel (b).}
\label{f3}
\end{figure*}
An example of these ideas is shown in Fig.~\ref{f3}.  We used very small patches from natural images as inputs, reshaping the intensities in nearby pixels into a $D$--component stimulus vector $\vek s$ where $D=10$.   To describe the neuron we chose a random symmetric matrix $\vek K$ to use as the kernel, and generated spikes when the stimulus energy, $\vek{s^\tr \cdot K\cdot s}$, crossed a threshold, as illustrated in Fig.~\ref{f3}a.  We fixed the mean rate of the spiking response such that the probability of a spike occurring in one bin of duration $dt$ is $0.1$, and we generated $\sim1000$ spikes. We then tried to extract the neuron's receptive field by starting with a random initial matrix $\vek Q$, and following the gradient of mutual information, as in Eq.~(\ref{grad_ascent}).  

We let the one parameter of the algorithm, $\gamma$, gradually decrease from a starting value of $0.5$ to $0.05$, in order to minimize the fluctuations around the true maximum of the information. Mutual information, the red trace in Fig.~\ref{f3}b, peaks at the $40^{\mathrm {th}}$ learning step and remains unchanged after that. The $3$ black dots in Fig.~\ref{f3}b correspond to the steps during the optimization when we extract and plot the initial guess, the intermediate and the optimal/maximally informative matrix $\vek Q$. It is interesting to note that the intermediate matrix appears completely different from the optimal $\vek Q$ even though the corresponding mutual information is relatively close to its maximum (a similar observation was made in the context of maximally informative dimensions \cite{mid}). 

In Fig.~\ref{f3}c the root--mean--square (RMS) reconstruction error ${\langle\left(\vek{K}-\vek{Q}\right)^2\rangle^{1/2}}$ is plotted as a function of the number of learning steps for a randomly initialized $\vek Q$ (solid line) and when $\vek Q$ is initialized to the spike--triggered covariance (STC) matrix (dashed line). RMS error at the start of the algorithm $\approx 1$ when the ``true" matrix $\vek K$ and the initial guess for $\vek Q$ are symmetric, random matrices, uncorrelated with each other, but is slightly lower when $\vek Q$ is initialized to the STC. This difference becomes smaller as the stimulus dimensionality $D$ increases or as the stimulus departs more strongly from Gaussianity. Both traces decrease, and stop changing once our estimate of the optimal $\vek Q$ matches $\vek K$. This occurs at the $40^{\mathrm{th}} $ step for the randomly initialized $\vek Q$ and slightly sooner ($30^{\mathrm{th}}$ step) when initialized to the STC. If we had fewer spikes, our estimate for the optimal $\vek Q$ could still match $\vek K$ adequately, but the actual RMS error in reconstruction might be higher. We explore such performance measures and data requirement issues next. 

\subsection{Performance measures and data requirements}

\begin{figure*}[!ht]
\begin{center}
\includegraphics[width=5.5in]{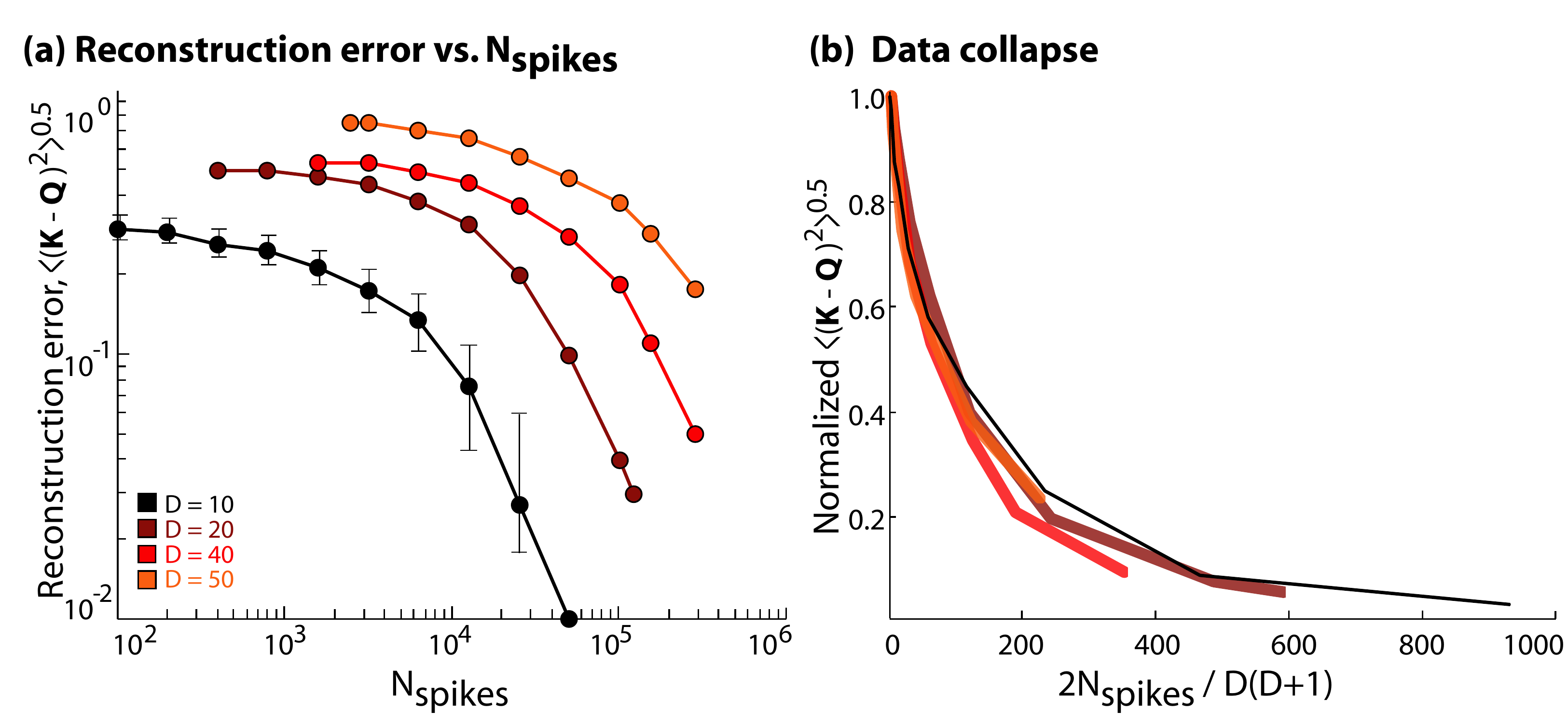}
\end{center}
\caption{
{\bf Data requirement and performance issues.} {\bf(a)} Reconstruction error $\langle(\vek K - \vek Q)^2\rangle^{1/2}$ is plotted as a function of number of spikes ($N_{\mathrm{spikes}}$) for matrices corresponding to stimuli of increasing dimensionality ($D = 10, 20, 40 \; \&\; 50$).  {\bf(b)} The traces for different values of $D$ collapse when the error is normalized by the (maximum) value at $N_{\mathrm{spikes}} = D(D+1)/2$ for each $D$, and plotted as a function of $2N_{\mathrm{spikes}}/(D(D+1))$.}
\label{f4}
\end{figure*}

We assume that spiking responses of the system are less frequent than periods of silence and therefore the accuracy of our reconstruction should depend of the number of spikes $N_{\mathrm{spikes}}$ produced by the model (or recorded data in the experimental context), independent of the actual threshold.   Further, we expect that when the stimulus dimensionality $D$ increases, the fact that there are more parameters needed to describe the kernel $\vek K$ means that performance will deteriorate unless we have access to more data.  To address these issues we explore model neurons as in Fig.~\ref{f3}, but systematically vary $D$ and $N_{\mathrm{spikes}}$.  Again we consider the most difficult case, with naturalistic stimuli and kernels that are random symmetric matrices.  The results are shown in Fig.~\ref{f4}.

Our intuition is that the number of spikes should scale with the number of free parameters in the model, $D(D+1)/2$, so we always start with $N_{\mathrm{spikes}} = D(D+1)/2$.  If we normalize the reconstruction errors by this initial error, and scale $N_{\mathrm{spikes}}$ by the number of free parameters, the data collapse, as shown in Fig.~\ref{f4}b.  Evidently accurate reconstructions of very large matrices requires very many spikes.  This suggests that it will be important to have some more constrained models, which we now explore.
\section{Constrained frameworks}

The most general stimulus energy is described by $D(D+1)/2$ parameters, and this quickly becomes large for high dimensional stimuli.  In many cases it is plausible that the matrix kernel of the stimulus energy has some simpler structure, which can be used to reduce the number of parameters.

One way to simplify the description is to use a matrix that has low rank.  If, for example, the rank of the matrix $\vek Q$ in Eq.~(\ref{stim_eng1}) is $p\leq D$, then we can find a set of orthogonal  vectors $\{v_i\}$ such that  
\begin{equation}
\vek{Q}=\sum_{i=1}^p \vek{v}_i\vek{v}_i^\tr .
\label{low_rank}
\end{equation}
In terms of these vectors, the stimulus energy is just $x =\sum_{i=1}^p (\vek v^\tr_i\cdot \vek s)^2$. 

The low rank approximation reminds us of the simpler, Euclidean notion of dimensionality reduction discussed above.  Thus, we could introduce variables $x_i = \vek v^\tr_i\cdot \mathbf s$ for $i=1, 2, \ldots , p$.  The response would then be approximated as depending on all of these variables, $r(x_1, x_2,..., x_p)$, as in  Eq.~(\ref{Kdims}).   In the stimulus energy approach, all of these multiple Euclidean projections are combined into $x = \sum_i x_i^2$, so that have a more constrained but potentially more tractable description.
When $\vek Q$ is written as $\vek{Q}=\sum_{i=1}^p \vek{v}_i\vek{v}_i^\tr$, the relevant gradient of information, analogous to Eq.~(\ref{IGrad}) is
\begin{widetext}
\begin{equation}
\nabla_{\vek{v_i}} I = 2\int dx\;P_{\vek Q}(x)\left[\left\langle\left.\vek s\, (\vek v_i^\tr \vek s) \right|x, \mathrm{spike}\right\rangle-\left\langle\left.\vek s\, (\vek v_i^\tr \vek s) \right|x\right\rangle\right]\frac{d}{dx}\left[\frac{P_{\vek Q}(x|\mathrm{spike})}{P_{\vek Q}(x)}\right],
\label{lrGrad}
\end{equation}
\end{widetext}
 and we can turn this into an algorithm for updating our estimates of the $\vek{v}_i$,
\begin{equation}
\vek v_i \rightarrow \vek v_i+\gamma\nabla_{\vek{v_i}} I .
\label{lowrnkascent}
\end{equation}
There is a free direction for the overall normalization of the matrix $\vek Q$ \cite{mid, rustV1} which makes the mutual information invariant to reparameterization of the quantities. We orthogonalize the vectors $\vek v_i$ during the learning procedure to make sure that they don't grow out of bound. 

Another way of constraining the kernel of the stimulus energy is to assume that it is smooth as we move from one stimulus dimension to the next.  Smooth matrices  can be expanded into weighted sums of basis functions, 
\begin{equation}
 \vek{Q} = \sum_{\mu=1}^M \alpha_\mu\;\vek{B}^{(\mu)} ,\label{basis}
 \end{equation}
and finding the optimal matrix then is equivalent to calculating the most informative $M$--dimensional vector of weights. 

The basis can be chosen so that systematically increasing the number of basis components $M$  allows the reconstruction of progressively finer features in $\vek{Q}$.  For example, we can consider $\{\vek{B}^{(\mu)}\}$ to be a set of Gaussian bumps tiling the $D\times D$ matrix $\vek{Q}$, and whose scale (standard deviation) is inversely proportional to $\sqrt{M}$. For $M\rightarrow D^2/2$ the basis matrix set becomes a complete basis, allowing every $\vek{Q}$ to be exactly represented by the vector of coefficients $\vek{\alpha}$. In  any matrix basis representation, the learning rule becomes,
\begin{equation}
\vek \alpha_\mu \rightarrow \vek \alpha_\mu+\gamma\sum_{{\rm i,j}=1}^M\frac{\partial\, I}{\partial \,\vek{Q}_{\rm ij}}\,\vek{B}_{\rm ij}^{(\mu)}.\label{basis_update}
\end{equation}
This is equivalent to taking projections of our general learning rule, Eq.~(\ref{grad_ascent}), onto the basis elements. 

\section{Results for model neurons}

\subsection{An auditory neuron}
 \begin{figure*}[!ht]
\begin{center}
\includegraphics[width=6.25in]{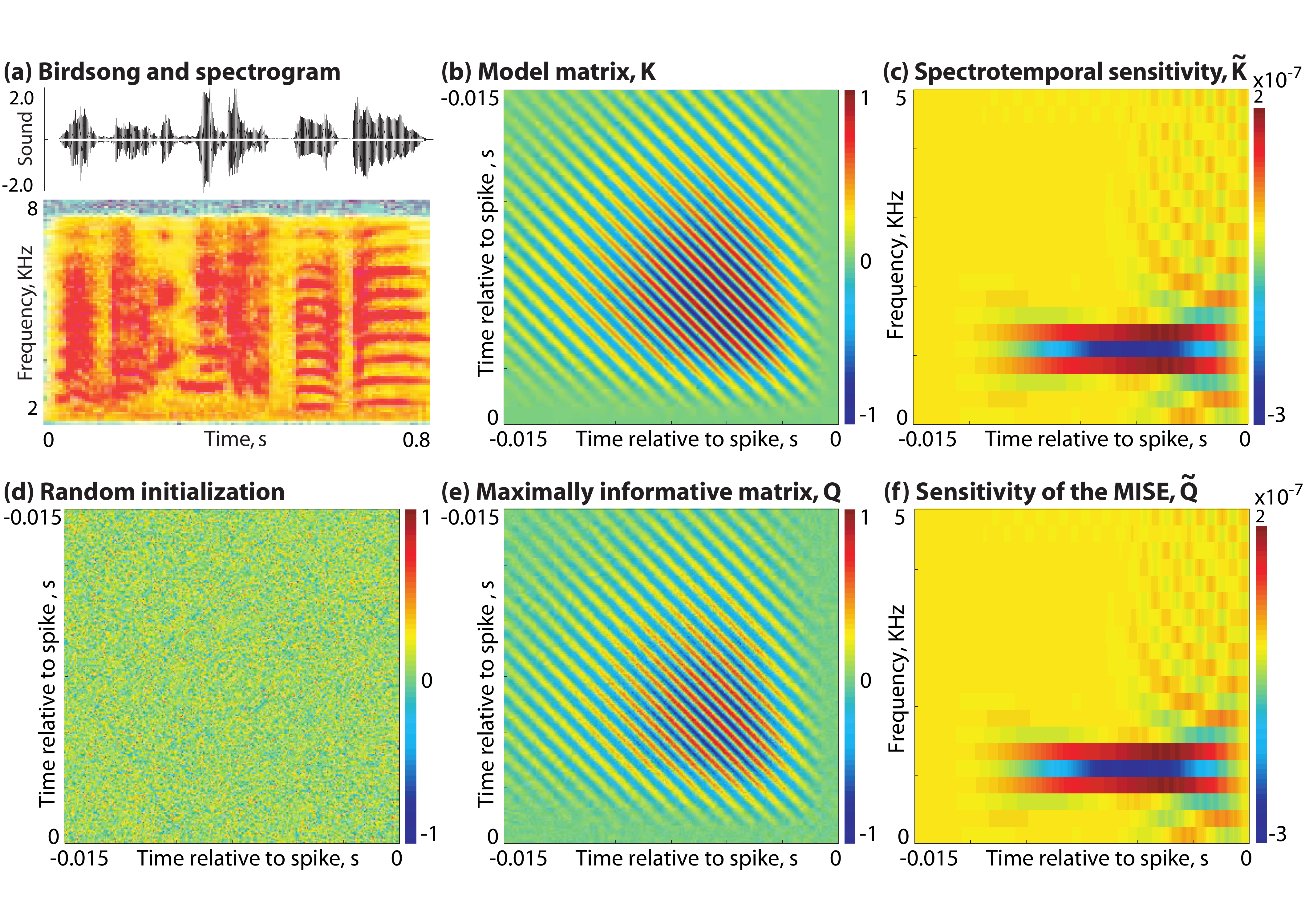}
\end{center}
\caption{
{\bf Analyzing the responses of the model auditory neuron to a bird song.}  {\bf(a)} The sound pressure wave of a zebra finch song used as stimulus to the model neuron is shown along with its spectrogram. The spectrogram of the song illustrates that $\vek s$ is highly structured, full of harmonic stacks and complex spectrotemporal motifs. {\bf(b)} The equivalent matrix $\vek K$, constructed from the two filters as described in Eq.~(\ref{aud_power}) is $300\times 300$ in size but has a relatively simple structure. {\bf(c)} Taking a Fourier transform over $t_2$ of $\vek K$ yields a spectrotemporal sensitivity matrix, $\vek {\tilde K}$ with a peak at approximately $1$KHz. {\bf(d)} The initial guess for $\vek Q$ is the random symmetric matrix plotted here. {\bf(e)} The optimal matrix $\vek Q$ that maximizes the mutual information between the spiking response of the model neuron and the $1D$ projection $x = \vek s^\tr\cdot \vek Q \cdot\vek s$ matches $\vek K$ well at the end of $100$ learning steps.  {\bf(f)} The spectrotemporal sensitivity $\vek {\tilde Q}$, corresponding to the maximally informative stimulus energy has the same response preferences as $\vek{\tilde K}$.}
\label{f5}
\end{figure*}
As a first example, we return to the model auditory neuron whose response properties from Eq's~(\ref{aud_power}) and (\ref{spk_prob}) were schematized in Fig.~\ref{f1}. Rather than studying its responses to white noise stimuli, however, we consider the responses to bird song, as shown in Fig.~\ref{f5}(a). We start with Eq.~(\ref{aud_power}) and see that it is equivalent to a stimulus energy with kernel $\vek K$ defined through
\begin{eqnarray}
p(t)&=& \int dt_1\int dt_2 \;s(t_1)\;\vek K(t - t_1, t-  t_2)\; s(t_2), \\
\vek K&=&\int d\tau f_2(\tau)f_1(t - \tau -t_1)f_1(t-\tau-t_2).
\label{equivalentK}
\end{eqnarray}
We used the same filters as we showed in Fig.~\ref{f1}(b) and (e) to construct $\vek K$, which is plotted in Fig.~\ref{f5}(b). We  can also look in a mixed time--frequency representation to generate a spectrotemporal ``sensitivity,'' $\vek {\tilde K}$, as follows:
\begin{equation}
\vek{\tilde K}(\omega, t) = \int d\tau \;\vek K\left( t+\frac{\tau}{2}, t-\frac{\tau}{2}\right)e^{+i\;\omega\tau}.
\label{sensitivity}
\end{equation}
Eq.~(\ref{sensitivity}) suggests that $\tilde{\vek K}$ is a function of $\omega$ and $t$, as seen in Fig.~\ref{f5}(c). $\tilde{\vek K}$ describes the selectivity of the model neuron to the stimulus, in this case the bird song.  We see that this neuron responds robustly to a sound with a frequency around $1$KHz with a temporal dependence dictated by the time constants of the $2$ filters that make up the neuron's receptive field matrix $\vek K$. This description has the flavor of a spectrotemporal receptive field (STRF), but in the usual implementations of the STRF idea a spectrogram representation is imposed onto the stimulus, fixing the shapes of the elementary bins in the time--frequency plane. Here, in contrast, Fourier transforms are in principle continuous, and we don't need to assume that the only relevant variable is stimulus power in each frequency band.

The natural stimuli we used to probe this model auditory neuron's receptive field came from recordings of zebra finch songs, modified into stimulus clips $\vek s$. The songs were interpolated down from their original sampling rate to retain the same discrete time steps ($dt=1/20000$s) that we use in Fig.~\ref{f1}. An example sound pressure wave of a song stimulus is plotted in Fig.~\ref{f5}(a). In the same panel, we emphasize the structural complexity of the stimulus in the spectrogram of the same song. The song spectrogram shows multiple motifs, significant temporal structure and several harmonic stacks, all of which point to the strong departure from Gaussianity of $\vek s$. 

This model neuron, as illustrated in Fig.~\ref{f1}, emitted spikes when the power in Eq.~(\ref{equivalentK}) exceeded a threshold. We set the threshold of firing so that the mean spike rate was $20\, {\rm s}^{-1}$. We presented $\sim 42$ minutes of bird song stimuli to the model neuron, collecting roughly $50,000$ spikes. 

We follow the gradient ascent procedure exactly as described in Eq.~(\ref{grad_ascent}).  Note that $\vek K$ in this model is a $300\times 300$ matrix, and we make no further assumptions about its structure; we start with a random initial condition (Fig.~\ref{f5}d).  The maximally informative matrix that we find is shown in Fig.~\ref{f5}e, and is in excellent agreement with the matrix $\vek K$; we can see this in the frequency domain as well, shown in Fig.~\ref{f5}f. Quantitatively, the RMS reconstruction error for this inference is less than  5\% of the maximum for any two uncorrelated random, symmetric matrices of the same size. 

\subsection{Complex cell in the visual cortex}

\begin{figure}[!ht]
\begin{center}
\includegraphics[width=3in]{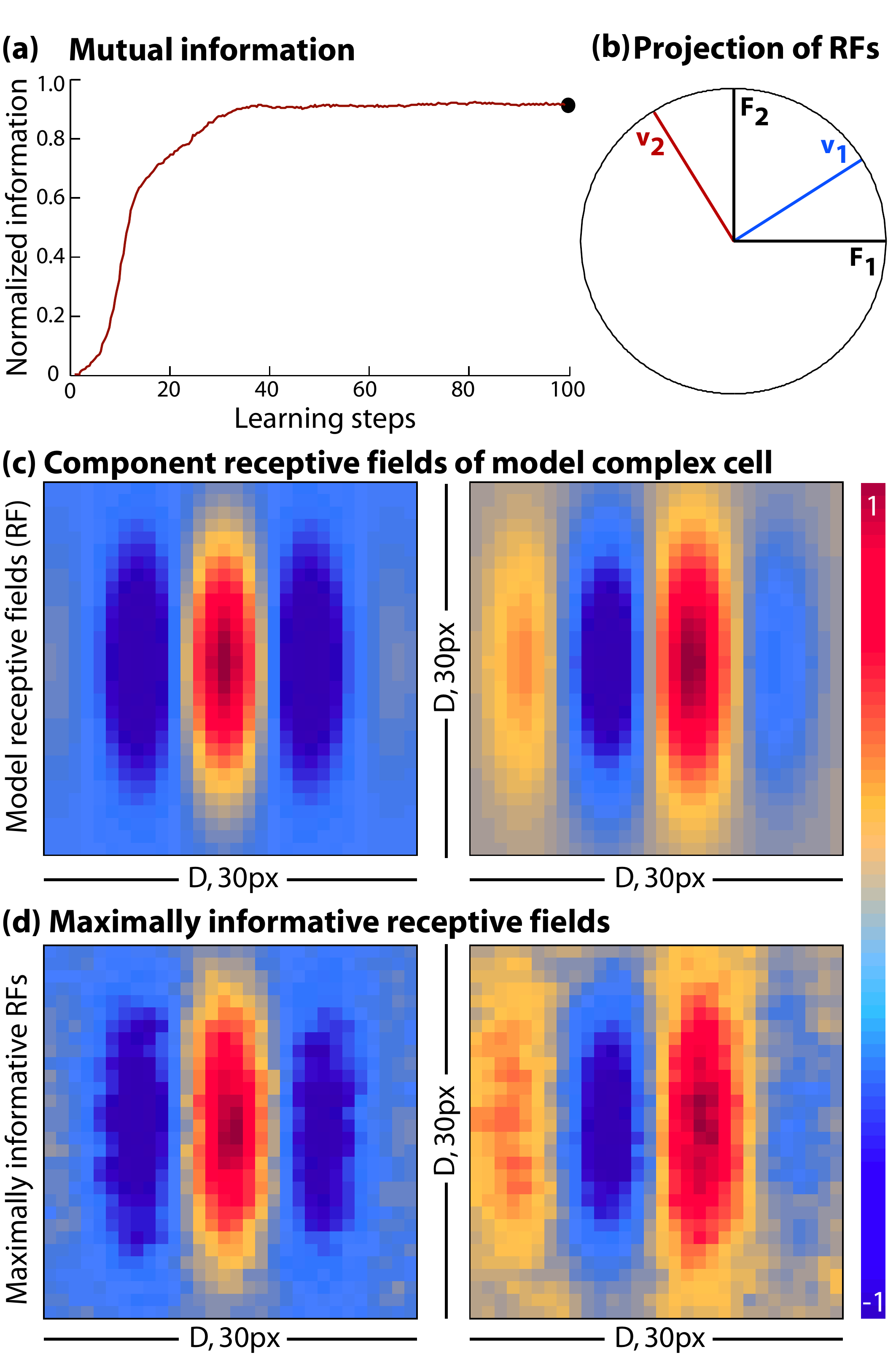}
\end{center}
\caption{
{\bf Receptive fields of a model complex cell and the reconstructed maximally informative pair.}  {\bf(a)} Information as a function of the number of learning steps peaks and then plateaus. The black dot is the point where the reconstructed receptive fields are shown in panel (d) below.  {\bf (b)}  Reconstructed vectors $\vek v_1$, $\vek v_2$ are rotated versions of the receptive fields $F_1$, $F_2$, but span the same linear subspace (all vectors are normalized to unit length). {\bf(c)} The receptive field of the model complex cell is given by the two linear filters in Eq.~(\ref{complexcell_power}): $F_1$ (left) and $F_2$ (right).  {\bf(d)} The reconstructed receptive fields at the $100^{\mathrm{th}}$ learning step (black dot in panel (a) above) with filters $\vek v_1$ (left) and $\vek v_2$ (right) rotated to best align with the $F_1$ -- $F_2$ pair.}
\label{f6}
\end{figure}

We consider the model complex cell described earlier in Eq.~(\ref{complexcell_power}), but allow integration over just one frame of a movie, so we don't need to describe the temporal filter.  We chose parameters $k  = 2\pi/3$, $\sigma_1 = 1.6$ and $\sigma_2 = 5$, with positions  measured in pixels.   The stimuli were 20,000 grayscale $30\times 30$ pixel image patches extracted from a calibrated natural image database \cite{pidb}.  Spikes were generated with a probability of a spike/bin of $0.1$ whenever the stimulus power $p$ exceeded a threshold.  We note that, in this model, the kernel is explicitly of rank 2, and so we followed the algorithm in Eq.~(\ref{lrGrad}). The results   are shown in Fig.~\ref{f6}. As expected, the best possible reconstruction is a vector pair $\vek v_1$, $\vek v_2$ that is equal to the pair $F_1$, $F_2$ up to a rotation.  Fig.~\ref{f6} shows that this is indeed the case for the reconstructed filters $\vek v_1, \vek v_2$, which otherwise match the true filters well. 
 
\begin{figure*}[!ht]
\begin{center}
\includegraphics[width=6.25in]{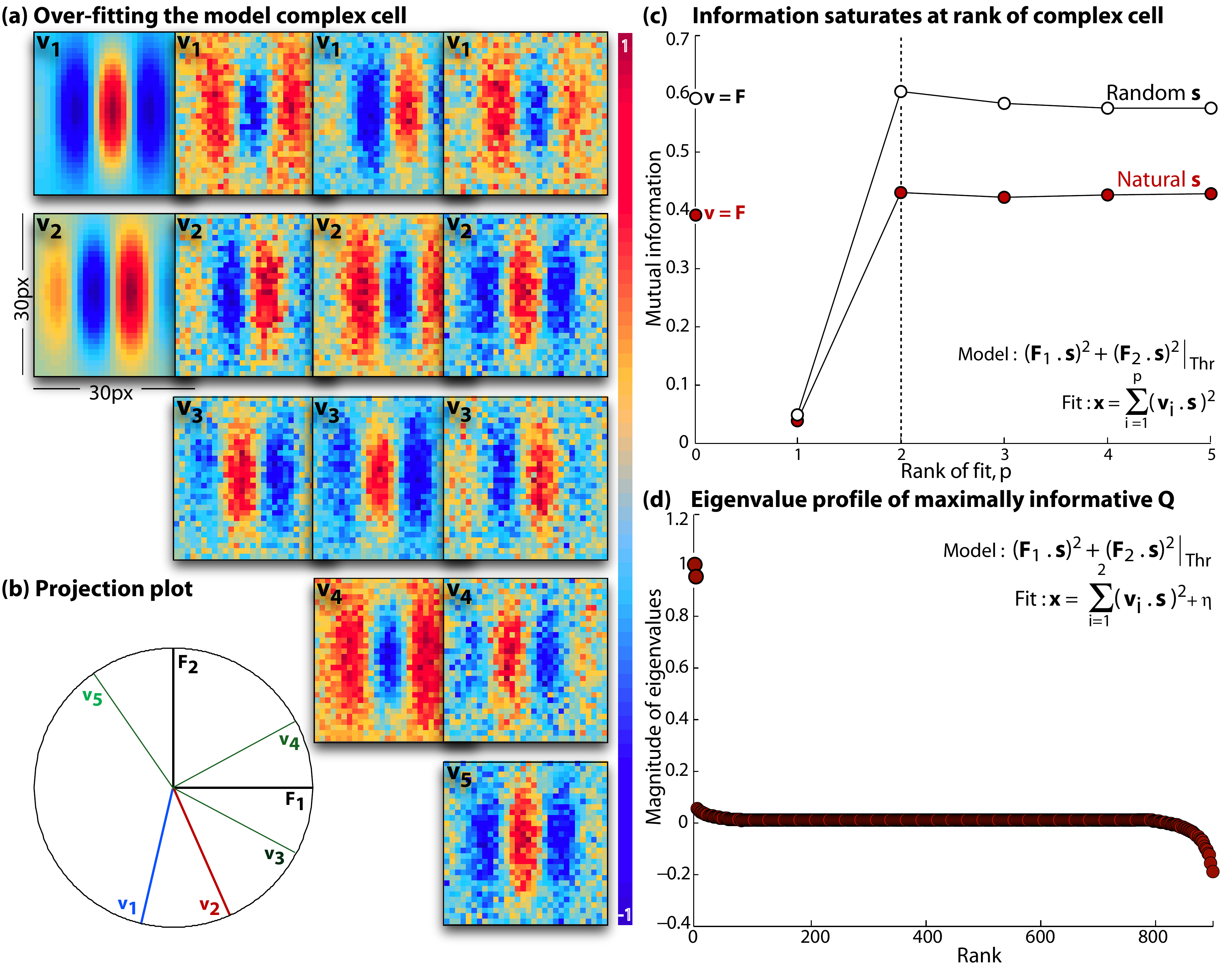}
\end{center}
\caption{
{\bf Over-fitting the model complex cell with matrices of successively increasing rank.} {\bf (a)} Receptive fields reconstructed after mutual information is maximized with matrices of rank $p = 2, 3, 4$ and $5$ (from left to right). {\bf (b)} The resulting vectors, $\vek v_1$ through $\vek v_5$, at the end of the information maximization are no longer orthogonal but project fully into a unit circle in the $F_1$--$F_2$ plane. {\bf (c)} Maximum mutual information as a function of the rank of fit, $p$, for random stimuli (open circles) or for the stimulus matrix generated using natural scenes (filled circles), peaks at the rank equal to that of the ``data" (rank $p = 2$ for the model complex cell), and remains unchanged as the rank of ${\vek Q}$ increases. {\bf (d)} Over-fitting the model matrix $\vek K$ with Gaussian noise does not add to the mutual information $I_{\vek Q}$ and the algorithm successfully finds a two dimensional solution. Eigenvalue profile of matrix $\vek Q = \sum_{i=1}^2\vek v_i\vek v^\tr_i + \eta$ where $\eta$ is a $30\times30$ sized-$\mathcal{N}(0, 1)$ after maximizing information with respect to the complex cell. Aside from two leading eigenvalues with magnitude $1$, the rest $\rightarrow 0$.}
\label{f7}
\end{figure*}

Suppose that the real neuron, as in our model, is described by a kernel of rank 2, but we don't know this and hence search for a kernel of higher rank. As shown in Fig.~\ref{f7}c, higher rank fits do not increase the information that we capture, either for random stimuli or for natural stimuli.  Interestingly, however, the ``extra'' components of our model are not driven to zero, but appear as (redundant) linear combinations of the two true underlying vectors, so that the algorithm still finds a genuinely two dimensional, albeit over--complete, solution.

The convergence of a full rank matrix $\vek {Q} = \sum_{i=1}^{2}\vek v_i\vek v_i^\tr+\eta$, where $\eta$ is Gaussian random noise of $\mathcal O(D)$, to the model matrix $\vek K$ can be determined by looking at the projections of the leading eigenvectors of the matrix ${\vek Q}$ at the end of the maximization algorithm. In this complex cell example where spikes are generated from a rank $2$ matrix, the two eigenvectors corresponding to the two leading eigenvalues for the fit, $\vek Q$ should be identical to what used to be $\vek v_1$ and $\vek v_2$ before. The remaining eigenvalues should be driven to be $0$, and this is indeed what we see in Fig.~\ref{f7}d. 
\subsection{Matrix basis formalism}

\begin{figure}[!ht]
\begin{center}
\includegraphics[width=3.2in]{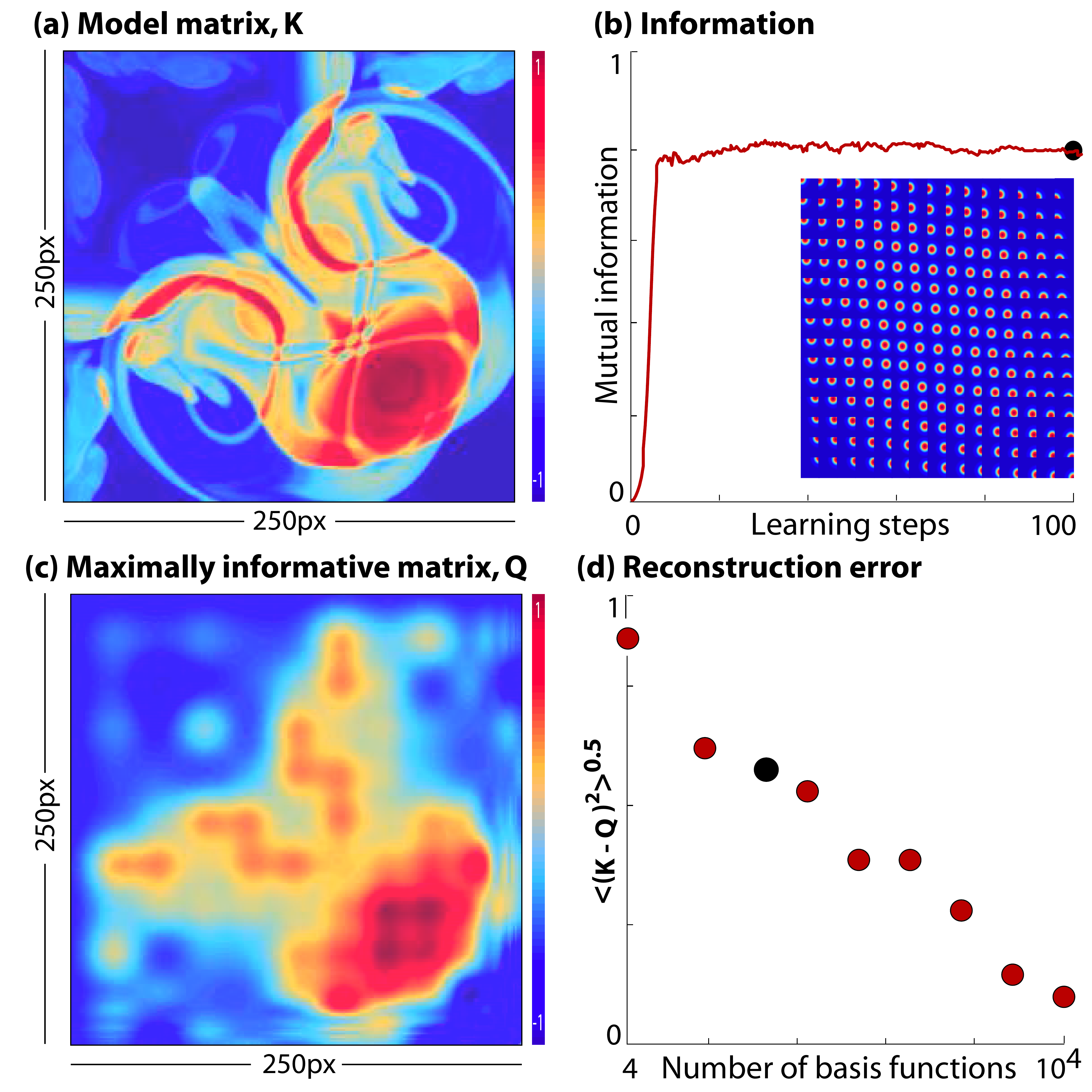}
\end{center}
\caption{{\bf Matrix basis formalism for constraining the number of free parameters.} {\bf (a)} A complex, structured full rank matrix ${\vek K}$ is generated by symmetrizing a $250\times 250$ pixel image of a fluid jet, and used as the ``true" kernel for our model  neuron. {\bf(b)} Mutual information increases with the number of learning steps, peaks at the $40^{\mathrm{th}}$ step and remains unchanged thereafter. Inset shows the collection of $225$ Gaussian matrix basis functions whose peaks densely tile the space of $\vek K$. A trial matrix is constructed as a linear sum (with coefficients $\{\alpha_\mu\}$) of the basis matrices, and information optimization is performed over  $\{\alpha_\mu\}$.  The black dot at  the $100^{\mathrm {th}}$ learning step is the point where $\vek Q$ is extracted. {\bf(c)} The reconstructed matrix kernel $\vek Q$ after maximizing mutual information using the $225$ basis coefficients making up the kernel is shown here. {\bf(d)} The RMS reconstruction error $\langle(\vek{K}-\vek{Q})^2\rangle^{1/2}$ decreases as the number of basis functions $M$ increases from $4$ to $10^4$. With enough data perfect reconstruction is possible as $M$ approaches the number of independent pixels in $\vek{K}$.}
\label{f8}
\end{figure}

We illustrate this simplification by making use of the matrix basis expansion from Eq.~(\ref{basis}) to infer a matrix $\vek{K}$ that is of arbitrarily high rank. For $\vek {K}$ we used a symmetrized $250\times250$ pixel image of a fluid jet as shown in Fig.~\ref{f8}a. While this is not an example of a receptive field from biology, it illustrates the validity of our approach even when the response has an atypical and complex dependence on the stimulus. Spikes were generated by thresholding the energy $\vek {s^\tr\cdot K\cdot s}$, and the same naturalistic visual stimulus ensemble was used as before. Gaussian basis matrices shown in the inset of Fig.~\ref{f8}b were used to represent the quadratic kernel, reducing the number of free parameters from $\sim 6\times10^4$ to $M=225$. We start the gradient ascent with a large $\gamma$ value of $1$ and progressively scale it down to $0.1$ near the end of the algorithm;  Fig.~\ref{f8}b shows the information plateauing in about 40 learning steps. The maximally informative quadratic kernel $\vek Q$ reconstructed from these $225$ basis coefficients is shown in Fig.~\ref{f8}c. Optimizing the $225$ basis functions captures the overall structure of the kernel but this can be improved to an almost perfect  reconstruction (at a pixel--by--pixel resolution) by increasing $M$, as shown in  Fig.~\ref{f8}d.
\subsection{Multiple matrices}

\begin{figure*}[!ht]
\begin{center}
\includegraphics[width=5.25in]{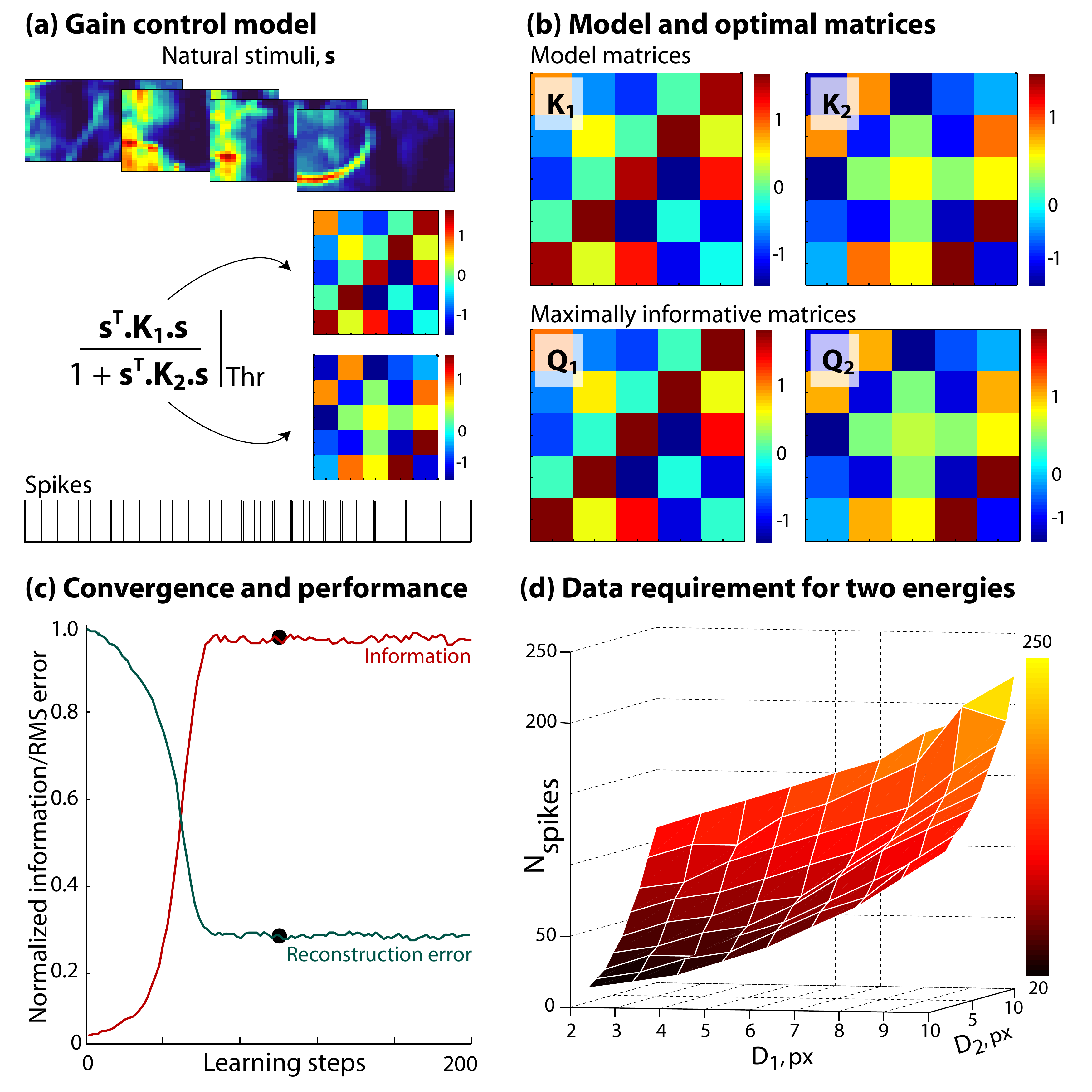}
\end{center}
\caption{
{\bf Inferring 2 MISE's simultaneously.}  {\bf(a)} Schematic of a model neuron with divisive gain control as described in the text. Natural images were reshaped and used as stimulus clips, $\vek s$. The ``true" matrices $\vek K_1$ and $\vek K_2$ were generated as random, symmetric $5\times 5$ matrices. {\bf(b)} The true ($\vek K_1$ and $\vek K_2$, top panel) and maximally informative matrices ($\vek Q_1$ and $\vek Q_2$, bottom panel) are plotted here. {\bf(c)} Mutual information (red) normalized by the maximum when $\vek K_1 = \vek Q_1$ and $\vek K_2 = \vek Q_2$, peaks at the $80^{\mathrm{th}}$ learning step and remains unchanged after. RMS reconstruction error (green), computed as $\langle\left[\vek{(K_1+K_2)}-\vek{(Q_1+Q_2)}\right]^2\rangle^{1/2}$ decays to a steady low at the same step. Using more spikes will decrease this further. {\bf(d)} The data requirement for simultaneously extracting $2$ matrices $\vek Q_1$ and $\vek Q_2$ to a precision of $30\%$ is shown in the shaded region. $N_{\mathrm{spikes}}$ here is proportional to the number of free parameters $\sim D_1^2 + D_2^2$, where $D_1$ and $D_2$ correspond to the stimulus dimensionality in the numerator and the denominator of $x$, respectively.}
\label{f9}
\end{figure*}

As a final test of our approach, we implemented the model neuron we described earlier in Eq.~(\ref{dual}), with stimulus dimensionality $D=5$; the two matrices, $\vek K_1$ and $\vek K_2$, are plotted in Fig.~\ref{f9}a, along with a schematic of the spikes produced when the nonlinearity $g(\cdot )$ is a threshold.  Again we constructed stimuli from nearby pixels of natural images, so that the distribution is strongly correlated and non--Gaussian; the threshold was set so that   the probability of a spike/bin to $0.3$, and we generated $\sim10,000$ spikes. We followed the algorithm in Eq's~(\ref{dualGrad}) and (\ref{dualGrad2}) to find the maximally informative matrices $\vek Q_1$ and $\vek Q_2$. The original and the optimal matrices are plotted in Fig.~\ref{f9}b, where see that for a model neuron with a randomly initialized duet of matrices, the optimal matrices agree with those of the model neuron. Mutual information in red, normalized by the maximum when $\vek K_1 = \vek Q_1$ and $\vek K_2 = \vek Q_2$, and RMS reconstruction error in green (calculated as $\langle\left[\vek{(K_1+K_2)}-\vek{(Q_1+Q_2)}\right]^2\rangle^{1/2}$) are plotted as a function of the learning steps in Fig.~\ref{f9}c. While convergence is definitely possible, our estimates of the maximally informative matrices are noisier than in the single matrix instances, even with a relatively large amount of data, as indicated in Fig.~\ref {f9}d.  We expect that realistic searches for multiple stimulus energies will require us to impose some simplifying structure on the underlying matrices.

\section{Discussion}

While the notion that neurons respond to multiple projections of the stimulus onto orthogonal filters is powerful, it has been difficult to develop a systematic framework to infer a neuron's response properties when there are more than two filters. To get around this limitation, we propose an alternative model in which the neural response is characterized by features that are quadratic functions of the stimulus. In other words, instead of being described by multiple linear filters, the selectivity of the neuron is described by a single quadratic kernel.  The choice of a quadratic form is motived by the fact that many neural phenomena previously studied in isolation can be viewed as instances of quadratic dependences on the stimulus. We presented a method for inferring maximally informative stimulus energies based on information maximization. We make no assumptions about how the quadratic projections onto the resulting matrices map onto patterns of spiking and silence in the neuron. This approach yields unbiased estimates for receptive fields for arbitrary ensembles of stimuli, but requires optimization in a possibly rugged information landscape. The methods we have presented should help elucidate how sensitivity to high--order statistical features of natural inputs arises in sensory cortical areas.
\section{Acknowledgments}
We thank AJ Doupe, RR de Ruyter van Steveninck, SR Sinha and G Tka\v{c}ik for helpful discussions and for providing the natural stimulus examples used here. We are also grateful to G Tka\v{c}ik and O Marre for providing critical scientific input during the course of this project. This work was supported in part by the Human Frontiers Science Program, by the Swartz Foundation, and by NSF Grant PHY--0957573.

\end{document}